\documentclass{article}

\usepackage{arxiv}

\usepackage[utf8]{inputenc} % allow utf-8 input
\usepackage[T1]{fontenc}    % use 8-bit T1 fonts
\usepackage{hyperref}       % hyperlinks
\usepackage{url}            % simple URL typesetting
\usepackage{booktabs}       % professional-quality tables
\usepackage{amsfonts}       % blackboard math symbols
\usepackage{nicefrac}       % compact symbols for 1/2, etc.
\usepackage{microtype}      % microtypography
\usepackage{lipsum}		% Can be removed after putting your text content
\usepackage{graphicx}
\usepackage{natbib}
\usepackage{doi}

\usepackage{caption}
\usepackage{subcaption}
\usepackage{graphicx}
\usepackage{float} 
\usepackage{amsmath,amsfonts,amssymb,amsthm}

\usepackage{xfrac}
\usepackage{xcolor}

\title{Multidimensional diffusion MRI methods with confined subdomains}

\author{
	{Deneb Boito$^{1,2}$, \, Cem Yolcu$^{1}$, \,  Evren \"Ozarslan$^{1,2}$}\\
	$^{1}$Department of Biomedical Engineering, Link\"oping University, Link\"oping, Sweden\\
	$^{2}$Center for Medical Image Science and Visualization, Link\"oping University, Link\"oping, Sweden \\
	
}

\date{}

\renewcommand{\headeright}{}
\renewcommand{\undertitle}{}
\renewcommand{\shorttitle}{md-dMRI with confined subdomains}

\begin{document}
	\maketitle
	
	\begin{abstract}
		Diffusion Magnetic Resonance Imaging (dMRI) is an imaging technique with exquisite sensitivity to the microstructural properties of heterogeneous media.  The conventionally adopted acquisition schemes involving single pulsed field gradients encode the random motion of water molecules into the NMR signal, however typically conflating the effects of different sources contributing to the water motion. Time-varying magnetic field gradients have recently been considered for disentangling such effects during the data encoding phase, opening to the possibility of adding specificity to the recovered information about the medium's microstructure. Such data is typically represented via a diffusion tensor distribution (DTD) model, thus assuming the existence of several non-exchanging compartments in each of which diffusion is unrestricted. In this work, we consider a model that takes confinement into account and possesses a diffusion time-dependence closer to that of restricted diffusion, to replace the free diffusion assumption in multidimensional diffusion MRI methods. We first demonstrate how the confinement tensor model captures the relevant signal modulations impressed by water diffusing in both free and closed spaces, for data simulated with a clinically feasible protocol involving time-varying magnetic field gradients. Then, we provide the basis for incorporating this model into two multidimensional dMRI methods, and attempt to recover a confinement tensor distribution (CTD) on a human brain dataset.
	\end{abstract}

% keywords can be removed

\let\thefootnote\relax\footnotetext{Corresponding Author: Evren \"Ozarslan, \texttt{evren.ozarslan@liu.se}}
	
\newpage
\section{Introduction}
	
	Diffusion Magnetic Resonance Imaging (dMRI) is a method used for investigating the microstructural organization of various heterogeneous media. This is achieved by sensitizing the MR signal to the random motion of water molecules inside the scanned substrate. To interpret and extract relevant information from the water motion, several models and signal representations have been developed. 
	
	At spatial resolutions achievable with current MR scanners, the scanned sample comprises several compartments within, outside, and possibly in between which diffusion is taking place. A general strategy for capturing this complexity without assuming any specific combination of compartments (see for example \citep{panagiotakiCompartmentModelsDiffusion2012} and \citep{jelescuDesignValidationDiffusion2017} for reviews of multi-compartment models for brain white matter), considers modeling the medium as a collection of isolated  pores, each represented by a diffusion tensor \citep{deswietPossibleSystematicErrors1996, jianNovelTensorDistribution2007}.  This approach leads to a diffusion tensor distribution (DTD), which could also be represented parametrically via normal \citep{BasserPajevic03} and Wishart \citep{jianNovelTensorDistribution2007} distributions as well as other related distributions \citep{Magdoom21, Shakya17Dagstuhl,Herberthson19}. Advances in diffusion encoding schemes \citep{coryApplicationsSpinTransport1990, wongOptimizedIsotropicDiffusion1995, Caprihan96, ozarslanGeneralFrameworkQuantify2009, westinMeasurementTensorsDiffusion2014} provided ways of disentangling confounding signal contributions, thus possibly enabling the extraction of relevant information about the medium's structure and composition via such modeling \citep{topgaardMultidimensionalDiffusionMRI2017}. However, it is rather paradoxical to have free diffusion within isolated compartments, as the cellular membranes have a strong effect on diffusion, making them the primary determinant of diffusion anisotropy \citep{Beaulieu94}. If this picture involving multiple isolated compartments is to be employed, it would be natural to represent the individual subdomains by accounting for confined diffusion within them \citep{Woessner63}.
	
	A viable alternative to the diffusion tensor representation of individual subdomains utilizes confinement tensors \citep{yolcuNMRSignalParticles2016} instead. In this case, the molecules are envisioned to be diffusing under the influence of an Hookean restoring force, i.e., according to the Ornstein-Uhlenbeck process \citep{uhlenbeckTheoryBrownianMotion1930}. Just like in restricted diffusion, the particle trajectories have limited extent, which has made the Ornstein-Uhlenbeck process a simple toy problem in earlier theoretical works on characterizing the influence of restricted diffusion on the NMR signal \citep{Stejskal65,ledoussalDecayNuclearMagnetization1992, mitraEffectsFiniteGradientPulse1995}. 
	
	Following a series of developments \citep{Ozarslan_ISMRM15_DICT,Zucchelli16ISBI}, the confinement tensor model has been noted to provide an alternative representation of diffusion anisotropy, very-well suited for studying heterogeneous media \citep{yolcuNMRSignalParticles2016, LiuOzarslan2019NBMreview, Afzali21review}. Furthermore, for NMR experiments involving long diffusion encoding pulses, the harmonic confinement becomes the effective model of restricted diffusion, giving rise to an approximately linear dependence of the effective stochastic force on the center-of-mass position of the particles during the application of the gradient pulses \citep{ozarslanEffectivePotentialMagnetic2017}. 
	
	Similarly to the better-known diffusion tensor model, the model proposed by \citet{yolcuNMRSignalParticles2016} captures the pore's geometry/anisotropy with a tensorial object, which can be visualized as an ellipsoid. However, the confinement tensor model offers an extra parameter to encode diffusivity. This parameter can either be a scalar bulk diffusivity, or another tensorial quantity. In either case, this represents the diffusivity when there is no impediment to the particles' motion, i.e., when the confinement value approaches $0$. Therefore, the confinement tensor model can accommodate both restricted and unrestricted diffusion. In a recent study, the orientationally-averaged signal was studied for confined diffusion measured via single- and double diffusion encoding measurements demonstrating that certain features of the NMR signal \citep{Mitra95,OzarslanJCP08} that cannot be predicted by diffusion tensors are reproduced by the confinement tensor model \citep{Yolcu2021AAFS}.
	
	These findings suggest the confinement tensor model as a plausible alternative for representing non-exchanging microscopic domains in multicompartment specimen models. In this work, we therefore propose to incorporate this model into the so-called multi-dimensional MRI methods \citep{topgaardMultidimensionalDiffusionMRI2017}. In particular, we replace the diffusion with the confinement tensor in Diffusion Tensor Distribution Imaging (DTDI) \citep{jianNovelTensorDistribution2007, topgaardDiffusionTensorDistribution2019}, and illustrate that the moments of the DTD estimated using Q-space trajectory Imaging (QTI) \citep{westinQspaceTrajectoryImaging2016} would have a different interpretation for confined diffusion. We start by assessing the capabilities of the confinement model in representing single pores on data simulated using a typical protocol involving general time-varying diffusion gradient fields \citep{szczepankiewiczLinearPlanarSpherical2019}, and then proceed with first attempts at recovering distributions of confinement tensors in a human brain dataset.

	\newpage\section{Background and Theory} 
	
	\subsection{Diffusion under a Hookean restoring force}
	
	In a diffusion NMR experiment, diffusing molecules acquire a phase shift depending on their trajectory $\mathbf{x}(t)$ and on the time-varying magnetic field gradient $\mathbf{G}(t)$. The signal from all molecules can be expressed as
	\begin{equation}
		\label{eq:averageNMR}
		E = \left\langle e^{-i\gamma \textstyle\int \mathrm{d}t \, \mathbf{x}(t) \cdot \mathbf{G}(t)} \right\rangle \ ,
	\end{equation}
	where $\gamma$ is the gyromagnetic ratio, and the averaging is performed over all particle trajectories. 
	
	For the case of diffusion under a Hookean restoring force, we shall denote by $\mathbf{C}$ the confinement tensor, which, upon multiplication by the Boltzmann constant $k_B$ and absolute temperature $T$, gives the tensorial force constant $\mathbf{f} = k_B T\mathbf{C}$ defining the Hookean potential $V$ through $V(\mathbf{r}) = \frac{1}{2} \mathbf{r}^\intercal  \mathbf{f} \mathbf{r}$. Furthermore, we denote the possibly anisotropic \textit{free} diffusion tensor by $\mathbf{D}$, and assume that $\mathbf{D}$ and $\mathbf{C}$ commute, i.e., they share the same  eigendirections. Finally, we introduce $\mathbf{\Omega} = \mathbf{D} \mathbf{C}$ for brevity. 
	
	Statistical quantities, such as the signal, can be calculated using the path weight
	\begin{equation}
		\mathrm{Pr}[\mathbf x()]  \propto \exp\left(  -\frac{1}{4}\int \mathrm{d}t \,  \left(\frac{\mathrm{d}\mathbf{x}}{\mathrm{d}t} + \mathbf{\Omega} \mathbf{x}(t)\right)^\intercal \mathbf{D}^{-1}  \left(\frac{\mathrm{d}\mathbf{x}}{\mathrm{d}t} + \mathbf{\Omega} \mathbf{x}(t)\right)\right) \ ,
	\end{equation} 
which represents the differential probability for a particle to follow the trajectory $\mathbf{x}()$ .
	The NMR signal in \eqref{eq:averageNMR} can thus be evaluated, up to a constant, through the path integral
	\begin{align}
		E \propto \int \mathcal{D}\mathbf x() \, \exp{\left( -\int \mathrm{d}t \left(\frac{1}{4} \left(\frac{\mathrm{d}\mathbf{x}}{\mathrm{d}t} + \mathbf{\Omega} \mathbf{x}(t)\right)^\intercal \mathbf{D}^{-1}  \left(\frac{\mathrm{d}\mathbf{x}}{\mathrm{d}t} + \mathbf{\Omega} \mathbf{x}(t)\right) +i\gamma \mathbf x(t) \cdot \mathbf G(t)\right)\right)} \ .
	\end{align}
	Thanks to stationarity, the time integration can be taken from $-\infty$ to $\infty$, in which case employing the substitutions
	\begin{subequations}
		\begin{align}
			\mathbf x(t) &= \int \frac{\mathrm{d}\omega}{2 \pi} \, e^{i\omega t} \, \mathbf{\hat x}(\omega) \\
			%\end{subequations}
			%and %$G(t)$ with
			%\begin{equation}
			\mathbf G(t) &= \int \frac{\mathrm{d}\omega}{2 \pi} \, e^{i\omega t} \, \mathbf{\hat G}(\omega)
		\end{align}
	\end{subequations}
	yields
	\begin{equation} \label{eq:E(w)}
		E = \exp\left(-\int \frac{\mathrm{d}\omega}{2 \pi} \, \hat{\mathbf{G}}^\dagger (\omega) \, \mathbf{K}(\omega) \,  \hat{\mathbf{G}}(\omega)\right)
	\end{equation}
	with
	\begin{equation}\label{eq:K(w)}
		\mathbf{K}(\omega) = 2\gamma^2 \mathbf{D}(\omega^2\mathbf{I} + \mathbf{\Omega}^2)^{-1} \ .
	\end{equation}
	Converting equation \ref{eq:E(w)} to the time domain yields
	\begin{equation}
		\label{eq:confNoImp}
		E = \exp\left(- \frac{\gamma^2}{2} \int \mathrm{d}t \int \mathrm{d}t' \, \mathbf{G}^\intercal(t')\, \mathbf{D}\, \mathbf{\Omega}^{-1}\, e^{-\mathbf{\Omega}|t-t'|}\, \mathbf{G}(t)\right) \ .
	\end{equation}

\newpage	\subsection{The confinement tensor model}
	
\begin{figure}
		\centering
		\includegraphics[width=0.96\linewidth]{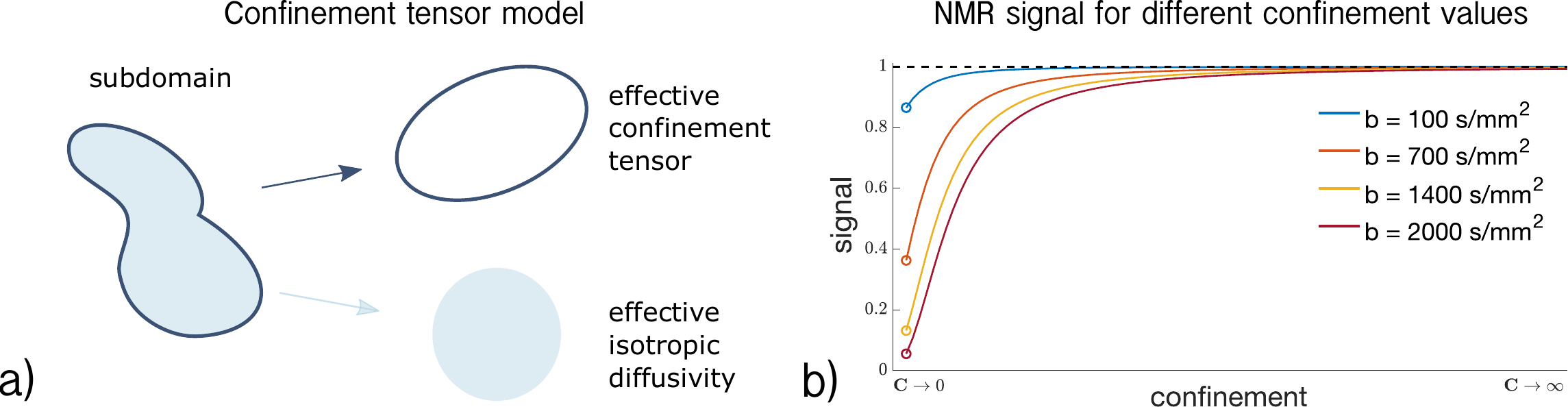}
		\caption{a) Confinement tensor model parameters. For a subdomain of generic shape, the geometry is captured by the effective confinement tensor, while the water diffusivity within is represented by an effective isotropic diffusivity. The tensor representing the shape is a second order tensor, while the effective diffusivity is a scalar.  b) Signal for different confinement values for linear tensor encoding experiments at different b-values. For $\mathbf{C} \rightarrow 0$ the signal converges to the free diffusion regime, where the value is determined as $\exp(-b \, D_0)$, with $b$ being the b-value \citep{LeBihan85}. This is represented by the colored circles in the plot. For $\mathbf{C} \rightarrow \infty$, the signal converges to $1$ indicating complete water immobility. } \label{fig:fig1}
	\end{figure}

	Figure \ref{fig:fig1}a shows the parameters of  the confinement tensor model \citep{yolcuNMRSignalParticles2016} and how they represent a generic pore. The shape of the subdomain is captured by an effective confinement tensor  $\mathbf{C}$ with units of inverse squared length, like in the case of diffusion under a Hookean force as described above. On the other hand, the rate of water diffusivity is captured by a scalar effective isotropic diffusivity $D_{\text{eff}}$.  
	
	While the expression given in  eq.\eqref{eq:confNoImp} is the natural way for defining the signal implied by the confinement tensor model, that is not optimal for the actual computation of the signal. To avoid potential numerical issues with the inversion of the $\mathbf{\Omega}$ tensor within the integral, we find more advantageous to use the equivalent expression given by \citep{yolcuNMRSignalParticles2016} for a gradient waveform applied between time points $0$ and $t_f$,
	\begin{equation}\label{eq:signal}
		E = \exp\bigg(- D_{\text{eff}} \int_{0}^{t_f} \mathrm{d}t \, |\mathbf{Q}(t)|^2\bigg) \, \exp \bigg(- \frac{D_{\text{eff}}}{2} \mathbf{Q}^\intercal(0) \, \mathbf{\Omega}^{-1} \, \mathbf{Q}(0)\bigg)
	\end{equation}
	with 
	\begin{equation}\label{eq:Q(t)}
		\mathbf{Q}(t) = \gamma \int_{t}^{t_f} \mathrm{d}t'  e ^{- \mathbf{\Omega}(t' - t)}\mathbf{G}(t') \ .
	\end{equation}
	
	Note that for $\mathbf{C} \rightarrow 0$, the signal in eq \eqref{eq:signal} reduces to the NMR signal expression for isotropic free diffusion (the proof is provided in \citep{yolcuNMRSignalParticles2016}), while for $\mathbf{C} \rightarrow \infty$, the signal  converges to $1$, indicating particles' immobility. Both these scenarios are shown in Figure \ref{fig:fig1}b, where the signal for confinement values in the range $[0, \infty)$ is shown.

	\vspace{20pt}\subsection{md - dMRI with confinement}
	The expressions derived thus far concern the MR signal for a single confinement tensor.  Here, we instead consider the case where a distribution of such tensors is collectively giving rise to the signal. In particular, we provide the signal expression for a confinement tensor distribution (CTD), which could be used for performing Confinement Tensor Distribution Imaging (CTDI), and discuss employing QTI for locally confined diffusion.

	\vspace{20pt}\subsubsection{Confinement tensor distribution (CTD)}
	\label{sec:CTD}
	The NMR signal expression for a distribution $\mathcal{P}(\mathbf{D})$ of diffusion tensors is given by \citep{jianNovelTensorDistribution2007}
	\begin{equation} \label{eq:dtd}
		S(\mathbf{b}) = S_0 \int \mathcal{P}(\mathbf{D}) \, e^{- \mathbf{b} : \mathbf{D} } \, \mathrm{d}\mathbf{D} \ ,
	\end{equation}
	where $\mathbf{b}$ is the measurement tensor \citep{Mattiello94}, $\mathbf{D}$ is the diffusion tensor, and $``:"$ indicates the generalized scalar product between tensors.  A similar expression can be introduced to include the confinement tensor model. Due to the extra parameter $D_{\text{eff}}$, the considered distribution becomes the joint distribution of effective confinement tensors and effective diffusivities $\mathcal{P}(\mathbf{C}, D_{\text{eff}})$. The signal expression for an experiment determined by a general time varying magnetic field gradient $\mathbf{G}(t)$  is 
	\begin{equation}\label{eq:ctd}
		S(\mathbf{G}(t)) = S_0 \int \mathcal{P}(\mathbf{C},D_{\text{eff}}) \,  E(\mathbf{G}(t), \mathbf{C}, D_{\text{eff}}) \, \mathrm{d}\mathbf{C} \, \mathrm{d}D_{\text{eff}} \ ,
	\end{equation}
   where $E(\mathbf{G}(t), \mathbf{C},D_{\text{eff}} )$ is as defined in equations \eqref{eq:confNoImp} or \eqref{eq:signal}. 
	Equations \eqref{eq:dtd} and \eqref{eq:ctd} can be considered to be generalizations of the Laplace transforms of  $\mathcal{P}(\mathbf{D})$ and $\mathcal{P}(\mathbf{C},D_{\text{eff}})$, respectively. Recovering the  $\mathcal{P}(\mathbf{D})$ (or $\mathcal{P}(\mathbf{C},D_{\text{eff}})$) from a series of measurements, i.e., numerically inverting the Laplace transform, is well known to be an ill-posed problem  \citep{galvosasMultidimensionalInverseLaplace2010,topgaardDiffusionTensorDistribution2019, reymbautAccuracyPrecisionStatistical2020}.

%\vspace{10pt}
\subsubsection{QTI for locally confined diffusion}
	The QTI technique exploits the sensitivity of the detected signal to the statistical moments of the structural parameters describing the specimen \citep{OzarslanNJP11}.  
	When the DTD model is employed for the latter, the high signal (low diffusion sensitivity) regime reveals the first few moments of the diffusivities\citep{westinQspaceTrajectoryImaging2016, Lasic14}.

For a DTD characterized by the distribution $P(\mathbf D)$ the signal decay in \eqref{eq:dtd} can be expressed by 
	\begin{align}%\label{eq:dtd}
		E(\mathbf{b}) = \langle e^{- b_{k\ell} \, D_{k\ell} } \rangle \ ,
	\end{align}
	where we employed the Einstein summation convention. At low diffusion sensitivity, the natural logarithm of the signal decay is approximated by the Maclaurin series of the above expression around $\mathbf b\approx \mathbf 0$, yielding
	\begin{align}
		\ln E(\mathbf{b}) \approx -b_{k\ell} \, \langle D_{k\ell} \rangle + \frac{1}{2} \, b_{k\ell} \, b_{mn} \, \langle D_{k\ell} D_{mn} \rangle_c \ ,
	\end{align}
	where the last quantity is the second cumulant, i.e., $\langle D_{k\ell} D_{mn} \rangle_c = \langle D_{k\ell} D_{mn} \rangle -  \langle D_{k\ell} \rangle  \langle D_{mn} \rangle $. We remind that the components of the measurement tensor are given through \citep{Mattiello94,Ozarslan15cmra}
	\begin{align}
		b_{k\ell} = \int_0^{t_f} \mathrm dt \int_0^t \mathrm dt' \int_0^t \mathrm dt'' G_k(t') \, G_\ell(t'') \ .
	\end{align}

	In the case of a CTD, the averaging takes the form of an integration over $D_\mathrm{eff}$ as well as $\mathbf{C}$; see \eqref{eq:ctd}. Applying the same procedure for the signal in frequency domain, \eqref{eq:E(w)}, yields
	\begin{align}
		\ln E(\hat{\mathbf{G}}(\omega) ) \approx & - \int \frac{\mathrm d\omega}{2\pi} \hat G_k(\omega) \, \hat G_\ell(\omega)  \, \langle K_{k\ell}(\omega) \rangle \\ \nonumber 
		& + \frac{1}{2} \int \frac{\mathrm d\omega}{2\pi} \int \frac{\mathrm d\omega'}{2\pi} \hat G_k(\omega) \, \hat G_\ell(\omega) \, \hat G_m(\omega') \, \hat G_n(\omega') \, \langle K_{k\ell}(\omega) \, K_{mn}(\omega') \rangle_c   \ ,
	\end{align}
	while the same is given in the time-domain by
	\begin{align}
		\ln E(\mathbf G(t) ) \approx & - \int \mathrm d t \, \int \mathrm d t' \, G_k(t) \, G_\ell(t') \, \langle H_{k\ell}(t,t') \rangle  \nonumber \\
		& + \frac{1}{2} \int \mathrm dt \int \mathrm dt' \int \mathrm dt'' \int \mathrm dt''' \, G_k(t) \, G_\ell(t') \, G_m(t'') \, G_n(t''') \, \langle H_{k\ell}(t,t') \, H_{mn}(t'',t''') \rangle_c    \ ,
	\end{align}
	where
	\begin{align}
		H_{k\ell}(t,t') = \frac{\gamma^2}{2} \mathbf C^{-1} e^{-\mathbf D \mathbf C |t'-t|}
	\end{align}
	since $\mathbf C$ and $\mathbf D$ commute.
	
		Note that, the shape of the waveform $\mathbf G(t)$ is inextricably linked to the signal in the CTD case. Furthermore, the interpretations of the signal decay rate are substantially different for the CTD and DTD assumptions. Thus, when QTI is performed, one can quantify only apparent moments of a DTD while the same analysis employing the CTD model would provide a more meaningful description of the low diffusion sensitivity regime. 
	
	\section{implementation}
	\label{sec:implementation}
	The confinement tensor model was implemented in Matlab (The Mathworks Inc, Natick, Massachussets) according to equations \eqref{eq:signal} and \eqref{eq:Q(t)}. Numerical integration was performed using the trapezoidal rule. The signal computation for a given confinement tensor and effective isotropic diffusivity was carried out in a reference frame in which $\mathbf{\Omega}$ is diagonal. This is achieved by rotating the measurement waveforms $\mathbf{G}(t)$ with the rotation matrix determined by the eigenvectors of $\mathbf{\Omega}$. This allows for computations to be carried out separately for each of the eigenvalues of  $\mathbf{\Omega}$ thanks to the separability of the model \citep{yolcuNMRSignalParticles2016}. This approach mitigates numerical issues that arise for small confinement values, in which case inverting $\mathbf{\Omega}$ may become problematic. Possible issues can be alleviated by considering a Taylor expansion for the second exponential factor in \eqref{eq:signal} to remove the dependency of that part of the signal on $\mathbf{\Omega}$. The derivation of the expression for computing the approximation of the signal using the Taylor expansion is provided in Appendix. %\ref{app:taylor}.
	
	To fit the confinement model to the data, we used the Matlab function \textit{lsqnonlin} with \text{Levenberg-Marquardt} as algorithm. The unknown estimated quantities consist of the signal without diffusion weighting $S_0$, the effective diffusivity $D_{\text{eff}}$, and the six unique components of the confinement tensor $\mathbf{C}$.  During the fitting, the tensor $\mathbf{\Omega}$ is replaced by its Cholesky factorization to ensure the positive semidefiniteness of the estimated confinement tensor \citep{koayUnifyingTheoreticalAlgorithmic2006}.  
	
	 To estimate the joint distribution of confinement tensors and effective isotropic diffusivities,  we adapted the existing technology implementing a Monte Carlo inversion of equation \ref{eq:dtd}, as detailed in \citep{reymbautAccuracyPrecisionStatistical2020} and retrieved from  \url{https://github.com/markus-nilsson/md-dmri}. As for the original implementation, we limit ourselves to the case of axisymmetric tensors. These can be represented using 4 parameters: the parallel and perpendicular confinement ($\text{C}_{para}$ and $\text{C}_{perp}$) capture the pore's geometry, while the other two define the pore orientation through the azimuthal ($\phi$) and polar ($\theta$) angles. Altogether, each pore is represented by 5 parameters ($\text{C}_{para}$, $\text{C}_{perp}$, $\phi$, $\theta$, and $D_{\text{eff}}$). While performing the inversion, these parameters are searched within the limits $8 \leq \text{log}_{10}(\text{C}_{para} / m^{-2}), \text{log}_{10}( \text{C}_{perp} / m^{-2}) \leq 12 $, $0.1 \leq (D_{\text{eff}} / \mu m^2 ms^{-1}) \leq 3.2  $, $0 \leq \phi \leq 2\pi$, $0 \leq cos(\theta) \leq 1$.
	 	For each voxel, the recovered $\mathcal{P}(\mathbf{C},D_{\text{eff}})$ can be visualized in 3D plots where  $\text{C}_{para}$ and $\text{C}_{perp}$ vary along the $x$ and $y$ axes, while $D_{\text{eff}}$ varies along the $z$ axis, respectively. The pore direction is encoded using the RGB color scale. We adopted the convention of displaying the color according to the main diffusion direction, not according to the direction of maximum confinement.
	
	\section{Results}
	\subsection{Signal for single compartments}
	In this section we investigate the capabilities of the confinement tensor model in capturing features of both free and restricted diffusion in data where the diffusion sensitization is achieved with general time varying magnetic field gradients. We employ a protocol featuring  217 measurements comprising Linear Planar, and Spherical Tensor Encoding (LTE, PTE, and STE respectively) \citep{szczepankiewiczLinearPlanarSpherical2019}.  We refer to this protocol as \textit{tensor encoding}. Signals for diffusion taking place in closed and open geometries were computed using this protocol. The pore shapes and the respective defining parameters were as follows: 
	
	\begin{itemize}
		\item[$\circ$] Free isotropic diffusion, $D_0 = 3  \, \mu m^2 / ms$
		\item[$\circ$] Stick compartment (uni-directional free diffusion), $D_0 = 2.5 \, \mu m^2 / ms$
		\item[$\circ$] Infinite cylinder,  $ r = 5 \,\mu m$, $D_0 = 3 \,\mu m^2 / ms$
		\item[$\circ$] Capped cylinder 1, $l =12 \, \mu m$, $r = 2 \, \mu m$, $D_0 = 2  \,\mu m^2 / ms$
		\item[$\circ$] Capped cylinder 2, $l = 10 \, \mu m$, $r = 1.5 \, \mu m$, $D_0 = 2.5 \, \mu m^2 / ms$,
		\item [$\circ$]Sphere, $r = 5\, \mu m$, $D_0 = 2\mu m^2 / ms$
	\end{itemize}
	where $D_0$ is the bulk diffusivity, $r$  is the radius, and $l$  is the length.  The diffusion tensor model was used to generate the signals for the free diffusion and the stick compartments, while the method described in \citep{ozarslanGeneralFrameworkQuantify2009} was used to generate the signals for the cylinders and the sphere. 
	
\begin{figure*} 
		\centering
		\includegraphics[width=0.96\linewidth]{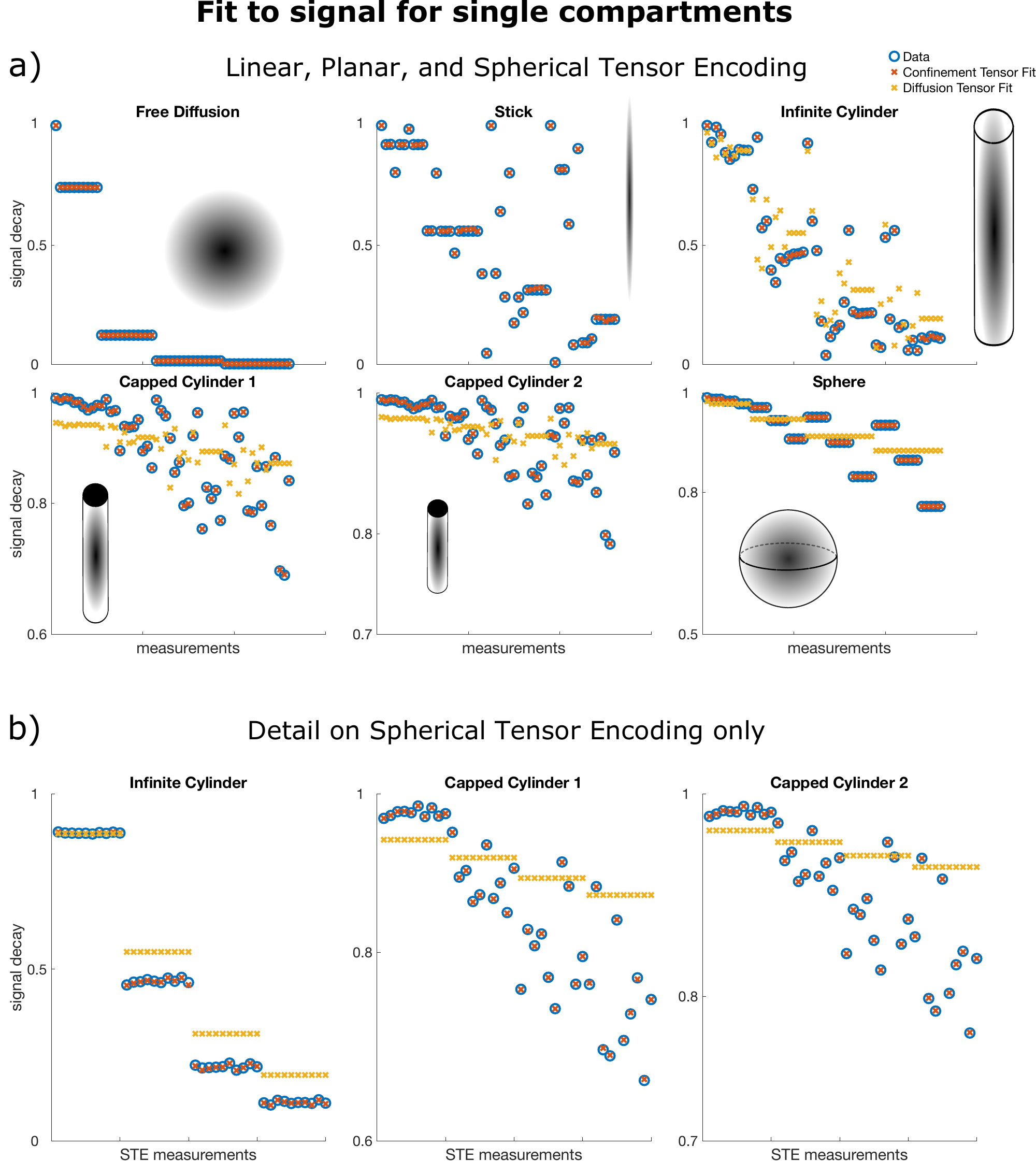}
		\caption{Confinement and diffusion tensor fit to signals arising from particles diffusing in free (\textit{Free Diffusion, Stick}) and restricted (\textit{Infinite Cylinder, Capped Cylinder 1, Capped Cylinder 2, Sphere}) geometries. In both panels a) and b) all the 217 data points were used to fit the model. For ease of interpretation, in panel a) only 52 random points out of the 217 are shown. In panel b) all the data points produced with Spherical Tensor Encoding are shown. Note how the confinement tensor model captures features of diffusion in both free and restricted scenarios. In particular, observe in panel b) how the signature of restricted diffusion imprinted on the signals produced with STE can be captured by the confinement tensor model while are inevitably missed by the diffusion tensor.} \label{fig:fig2}
	\end{figure*}

	Figure \ref{fig:fig2}a shows the results obtained by fitting the confinement tensor model and the diffusion tensor to the simulated signals. Only a subset of the measurements is shown for easier visual inspection. For the considered protocol, the confinement tensor model seems to capture well the features of both free and restricted diffusion, suggesting that the model given in equation \ref{eq:signal} has sufficient degrees of freedom and there is no need for employing a tensorial diffusivity.  Note that as illustrated for the 1D problem of restricted diffusion between two parallel plates \citep{ozarslanEffectivePotentialMagnetic2017},  in the ideal scenario involving only very long pulses and simple geometries, one would expect the measurement to be sensitive only to the product of squared confinement and diffusivity, in which case there is no need to employ an effective diffusivity in the model. However, in practice there is such sensitivity, and the relationship between the parameters of the model ($D_{\text{eff}}$, and $\mathbf{C}$) and those of the geometry requires further investigations. What is remarkable however is that having only one additional parameter ($D_{\text{eff}}$) offers sufficient complexity to capture the information in the signal for the considered acquisition scenario.
	 This is also evident in Figure \ref{fig:fig2}b, which better illustrates how the assumption of free diffusion fails \citep{deswietPossibleSystematicErrors1996, jespersenEffectsNongaussianDiffusion2019} while the confinement tensor model fully captures the signal modulations  due to restricted diffusion probed by STE measurements.
	 
	The recovered values of the effective diffusivity $D_{\text{eff}}$ coincided with the bulk diffusivity $D_0$ for the stick and free water compartments. For the compartments in which diffusion is fully restricted, the estimated values were lower than the nominal $D_0$. Respectively, $1.65 \, \mu m^2/ms$ for Capped Cylinder 1, $  2.0 \, \mu m^2/ms$ for Capped Cylinder 2, and $ 1.7 \, \mu m^2/ms$ for Sphere.
	
	\vspace{20pt}\subsection{Signal for a distribution of compartments}
	\label{sec:CTDDiB}	
\begin{figure}
		\centering
		\includegraphics[width=0.96\linewidth]{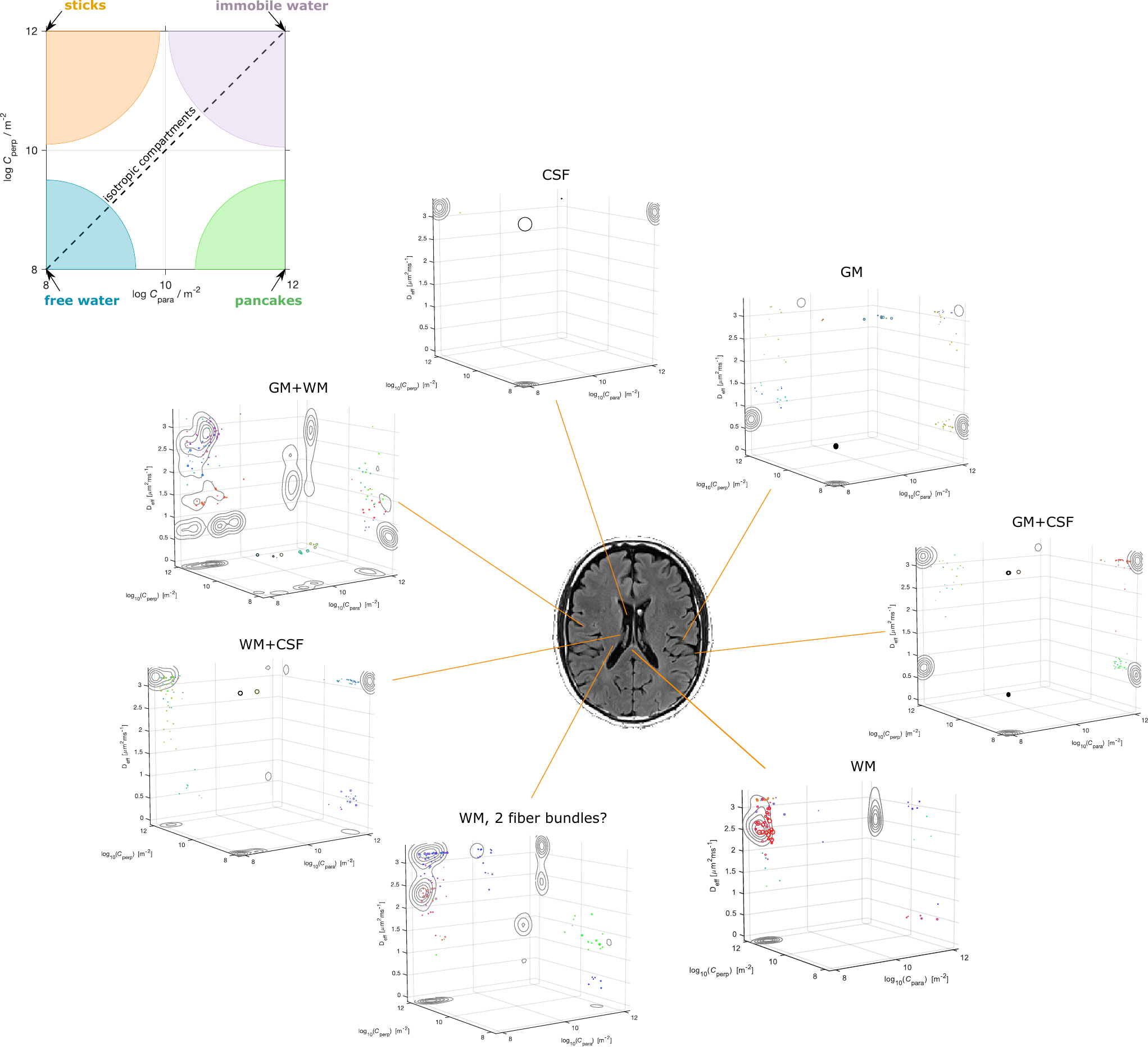}
		\caption{Example results of inverting equation \ref{eq:ctd} on real data collected with LTE, PTE, and STE waveforms \citep{szczepankiewiczLinearPlanarSpherical2019}. The panel on top shows a bird's eye view on the geometry plane, and how to associate the location of points to a shape according to the $\text{C}_{para}$ and $\text{C}_{perp}$ coordinates. The contours in the 3D plot show the projections of the pores' clusters onto the various 2D planes.} \label{fig:fig3}
	\end{figure}

	Illustrative results obtainable with the Monte Carlo inversion method described in section \ref{sec:implementation} were produced on a publicly available brain dataset \citep{szczepankiewiczLinearPlanarSpherical2019} collected via the \textit{tensor encoding} protocol used in the simulations. Figure \ref{fig:fig3} shows Monte Carlo inversion results on a few selected voxels on the bottom, and a bird's-eye view on the plane displaying the geometric information about the tensors in the distribution on the top. In this last, the colored areas indicate what shape each tensor would have for different values of  $\text{C}_{para}$ and $\text{C}_{perp}$. Pores with isotropic geometries are found along the diagonal where $\text{C}_{para} = \text{C}_{perp}$. Free isotropic diffusion is found for low $\text{C}_{para} = \text{C}_{perp}$, while confined isotropic diffusion is found for high $\text{C}_{para} = \text{C}_{perp}$. Stick-like pores are located at the $\text{C}_{para} \ll \text{C}_{perp}$ corner, while pancake-like pores are found at the $\text{C}_{para} \gg \text{C}_{perp}$ corner. 
	
	In the same spirit of what was shown in \citep{topgaardMultipleDimensionsRandom2019} for the DTD model, the 3D plots in Figure \ref{fig:fig3} show what the  $\mathcal{P}(\mathbf{C},D_{\text{eff}})$ for voxels containing either single or multiple types of brain tissues, as obtained from the data, \textit{could} be. For example, voxels containing pure CSF would have a $\mathcal{P}(\mathbf{C},D_{\text{eff}})$ of only free diffusion geometry with $D_{\text{eff}}\approx 3.1 \mu m^2 / ms$. Pure white matter voxels would only contain collections of stick-like geometries (see the voxel from the Corpus Callosum), while, for the considered dataset, gray matter \textit{could} contain isotropic free water at lower $D_{\text{eff}}$ compared to free water. Voxels with mixed tissue types could build their  $\mathcal{P}(\mathbf{C},D_{\text{eff}})$ based on those from single tissues. Note how having separate components encoding for the pore geometry ($\mathbf{C}$) and the water diffusivity ($D_{\text{eff}}$) allows for clearly identifying scenarios where pores could have the same shape, but different water mobility. See for example the voxel ``WM, 2 fiber bundles?'', where the distribution seems to suggest the presence of two differently oriented fiber bundles, which can be teased apart also by looking at their water diffusivity. The same specificity could not be achieved by only considering a distribution of diffusion tensors, where the information about the pore geometry is inextricably entangled to that of  water diffusivity.

	Note however that all what is presented and discussed in this section are simply initial conjectures, which may very well be the results of falling into the temptation of over-interpreting the outcomes of the DTD or CTD estimation. As shown by \citep{reymbautAccuracyPrecisionStatistical2020}, inverting equation \ref{eq:dtd}, and by extension equation\ref{eq:ctd}, is already challenging even at infinite SNR. The situation worsens in real data where the validity of the solutions proves to be very sensitive to the presence of noise. Moreover, as we present and discuss later, the results, and their interpretation, strongly depend upon the adopted acquisition scheme. For example, we could expect to find sphere-like compartments in gray matter in data encoded with different diffusion times and higher diffusion sensitivity, possibly indicating that a relevant fraction of the signal is due to cell bodies. Moreover, while the considered protocol (and data) should encode sufficient information for accurately recovering the pores' geometry, other waveforms could prove beneficial to study the time-dependence of the diffusion process, augmenting the reliability of the $D_{\text{eff}}$ dimension.

	\section{Discussion}

	The results in Figure \ref{fig:fig2} illustrate how well the confinement model captures the features of  both free and restricted diffusion, for data simulated with a clinically-feasible protocol including typical time-varying magnetic field gradients.  The signal's modulation due to restrictions is, under the considered experimental set-up, fully described by studying the problem of diffusion occurring in a potential landscape. This shows that the considered approach retains the right number of degrees of freedom to characterize diffusion processes within individual compartments. This  finding is consistent with what was reported by \citep{ozarslanEffectivePotentialMagnetic2017} for experiments involving long duration pulses, and supports the idea of  adopting the confinement tensor for representing isolated pores in multicompartment models.
	
	Having a single model covering both restricted and unrestricted diffusion in different geometries could be advantageous when defining multi-compartmental models based on such shapes as building blocks. Biophysical models, such as the composite hindered and restricted model of diffusion (CHARMED) \citep{Assaf05} and neurite orientation dispersion and density imaging (NODDI) \citep{Zhang12} strive for modeling specifically the neural tissue, therefore are not suitable for different tissues and other heterogeneous media. The confinement tensor representation of each compartment could be integrated into such models and could provide a convenient means for accounting for restricted diffusion. On the other hand, the confinement tensor distribution model is far more general than such specific models as one would not need to make \emph{a priori} assumptions on the specimen composition, apart from limiting its representation to numerous non-exchanging and possibly confined domains. The results in Figure \ref{fig:fig3} exemplified the specificity achievable by modeling a specimen with a joint distribution of confinement tensors-isotropic effective diffusivities. Other information about the water pools, such as $T_1$ and $T_2$ relaxations \citep{Callaghan07, topgaardMultipleDimensionsRandom2019}, could be added to increase specificity to the tissue heterogeneity. Similarly to what was presented there, the confinement tensor model could also be considered for diffusion - relaxation studies \citep{topgaardMultipleDimensionsRandom2019, benjaminiMultidimensionalCorrelationMRI2020, slatorCombinedDiffusionRelaxometry2021, AlmeidaMartins21HBM, Martin21}. 
	
	Note that on the specific dataset used in this work we did not observe striking modulation in the signal for isotropic measurements at constant  b-value. This could be explained by the experiments not being sensitive to finite-sized anisotropic restrictions, i.e., axons could effectively be pictured as sticks. Under these conditions, the fit to signal for both the DTD and CTD would yield very close results. Having two fundamentally different models exhibiting good fits to the data suggests that the data is possibly not descriptive enough. Another factor contributing to equal performance could be found in both DTD and CTD being overly-parameterized, thus effectively having the capabilities to fit the data equally successfully. This should not however be interpreted as both models being acceptable and providing informative results. In addition, based on the results in Figure \ref{fig:fig2}, we expect the CTD to provide meaningful information on data where restrictions have imprinted a clearer signature.
	
	We would also care to iterate once more on the limitation of performing CTDI (or DTDI) using the technology implemented in  \url{https://github.com/markus-nilsson/md-dmri}, due to the mathematically ill-posed problem that is being attempted. Different $\mathcal{P}(\mathbf{C},D_{\text{eff}})$  in \ref{eq:ctd} will represent the signal equally well, thus possibly leading to wrong interpretations of the microstructural characteristics of the scanned specimen. A similar issue is referred to as the ``degeneracy problem" \citep{jelescuDegeneracyModelParameter2016} in recovering the brain microstructure.  Multi-compartment models present flat fitting landscapes with multiple local minima located in different parts of the parameter space, each of which providing equally sound biological explanation for the signal. One approach to alleviate the problem involves including additional measurements, e.g, diffusion measurements having different temporal profiles, with the goal of disentangling the contribution of different parameters to the model interpretation \citep{ coelhoResolvingDegeneracyDiffusion2019}.
	
	When attempting at recovering the joint distribution of confinement tensors and isotropic effective diffusivities, we found from simulations that the pores' geometry can be obtained relatively faithfully using the \textit{tensor encoding} protocol only. However, determining the pore diffusivity relatively to the restriction size from data encoded exclusively in such manner seems to be more challenging. We provide examples of this in Figure \ref{fig:fig4}. In Figure \ref{fig:fig4}a we show a modified version of the \textit{tensor encoding} protocol, where half of the waveforms are replaced with Trapezoidal-Cosine Oscillating Gradient Spin-Echo (TC-OGSE). We refer to this protocol as \textit{mixed}. The goal is to achieve higher sensitivity to molecules' diffusivity within restrictions by using waveforms with well defined encoding frequency, and by matching the frequency of such waveform to that of the diffusion process \citep{Stepisnik93, lundellMultidimensionalDiffusionMRI2019,drobnjakOptimisingOscillatingWaveformshape2013}. Retaining part of the original protocol should ensure accurate pore geometry estimation. In Figures \ref{fig:fig4}b and \ref{fig:fig4}c we show the results obtained for a simple scenario where the specimen consists of two pools of water in which molecules are diffusing at different rates. When the data are simulated with only LTE, PTE and STE, it is possible to accurately recover the expected pore shapes but not the water diffusivity. Conversely, by introducing oscillating gradients, the diffusivity estimates, although still uncertain, converge to the correct values. 
	
	In Figure \ref{fig:fig4}d we show the results on a more complex substrate consisting of a stick compartment, an extra-axonal compartment, a sphere compartment and a compartment with free diffusion. As for the simple scenario described above, the estimation of the pores' diffusivity, in particular one of the sphere compartments, improves when the mixed protocol is used. This corroborates the idea of including measurements not only exploring different shapes of the encoding tensors, but also probing different frequencies \citep{lundellMultidimensionalDiffusionMRI2019}. The results in Figure \ref{fig:fig4}d also exemplify the uncertainty around inverting the Laplace transform, even for infinite SNR.

	\section{Conclusion}
	In this work we incorporated the confinement tensor model for  individual subdomains of heterogeneous media into multidimensional diffusion MRI frameworks.  We demonstrated how considering Brownian motion as taking place under the influence of a Hookean potential provides sufficient degrees of freedom to capture the signal modulations arising from water diffusing in restricted geometries. We argued that the confinement tensor distribution (CTD) model is a viable alternative to the diffusion tensor distribution model as CTD relies on the effective model of restricted diffusion, which makes it more consistent with the multicompartmental organization of complex tissue when examined via commonly performed diffusion MRI measurements. Despite its challenges, incorporating this model into multidimensional diffusion MRI methods could provide new insights regarding the structural composition of complex media.

	\begin{figure}
		\centering
		\includegraphics[width=0.98\linewidth]{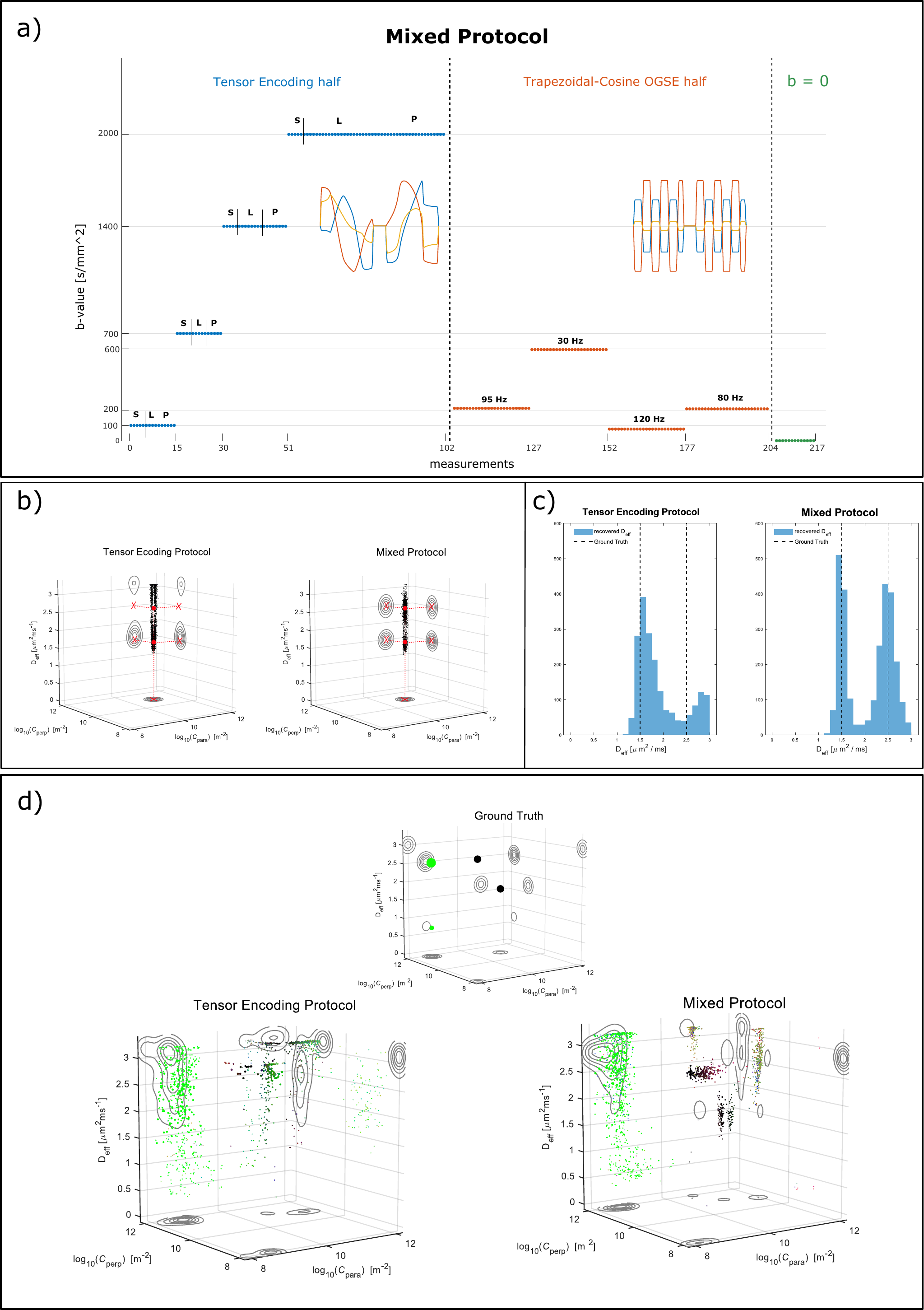}
	\end{figure}
	\begin{figure}
	\centering
	\newpage\caption{a) Layout of the \textit{mixed} protocol. $102$  measurements consist of LTE, PTE, and STE waveforms as defined in the \textit{tensor encoding} protocol, but with their directions redistributed over the sphere to achieve more uniform directional sampling. Another $102$ measurements consist of Trapezoidal Cosine Oscillating Gradient Spin Echo waveforms oscillating at $4$ different frequencies. The last $13$ measurements had null diffusion gradient strength. b) Results of inverting equation \ref{eq:ctd} for a simple system consisting of two equally weighted water pools with identical geometries but different water diffusivities. Left, data encoded with the \textit{tensor encoding} protocol. Right, data encoded with the \textit{mixed} protocol. The red filled dots depict the ground truth values in the 3D plot, while the red crosses show the ground truth values for the projections onto the various planes. c) The distributions of $D_{\text{eff}}$ obtained in b). d) Results inverting equation \ref{eq:ctd} for a substrate comprising a free water compartment, a sphere compartment, a stick compartment, and an intra-axonal compartment, with the following weights in the distribution: $0.25, 0.25, 0.4, 0.1$. The substrate ground truth is visualized in the small 3D plot on top. On the left, data encoded with the \textit{tensor encoding} protocol. On the right, data encoded with the \textit{mixed} protocol}.
	\label{fig:fig4}
\end{figure}

\section*{Author Contributions}

DB: Methodology, Software, Validation, Formal analysis, Investigation, Writing - Original Draft, Visualization.\\
CY: Methodology, Theory, Writing - Review \& Editing. \\
EÖ: Conceptualization, Methodology, Writing - Review \& Editing, Supervision, Project administration, Funding acquisition. 

\section*{Funding}
This project was financially supported by Linköping University Center for Industrial Information Technology (CENIIT), Analytic Imaging Diagnostics Arena (AIDA), and Sweden's Innovation Agency (VINNOVA) ASSIST.

\section*{Acknowledgments}
The authors thank Magnus Herberthson for stimulating discussions. 

\newpage
\bibliographystyle{unsrtnat}

\newpage\appendix
%------------------------------------------------------------------------
\section*{Appendix: Numerical approximation of the signal in $\mathbf{C} \rightarrow 0$ regime}\label{app:taylor}
The matrix inversion in the second exponential  in equation \eqref{eq:signal} can become numerically unstable for $\mathbf{C} \rightarrow 0$. When working in the coordinate system determined by the eigenvectors of $\mathbf{C}$, the matrix $\mathbf{\Omega}$  is diagonal
\begin{equation}
	\mathbf{\Omega} = \begin{bmatrix}
		\lambda _1 & 0 & 0 \\
		0 & \lambda _2 & 0 \\
		0 & 0 & \lambda _3
		\end{bmatrix} \ .
	\end{equation}
We take $\vec{\mathbf{v}}_1$, $\vec{\mathbf{v}}_2$, and $\vec{\mathbf{v}}_3$ to be the eigenvectors of $\mathbf{C}$, defining the new coordinate system for the experiment. The matrix $\mathbf R$ having $\vec{\mathbf{v}}_1$, $\vec{\mathbf{v}}_2$, and $\vec{\mathbf{v}}_3$  as columns, can be used to determine the gradient waveforms used to collect the data in the new coordinate system through
\begin{equation}
	\mathbf{G}'(t) = [G'_1(t), \, G'_2(t), \, G'_3(t)]^\intercal = \mathbf R^\intercal \,  \mathbf{G}(t) \ .
\end{equation}
Then, the signal contribution from the second exponential in equation \eqref{eq:signal} can be written as
\begin{multline}
	\label{eq:eugly}
	\text{exp} \left(- \frac{D_{\text{eff}}}{2} 
	\begin{bmatrix}
		Q_1(0) & Q_2(0) & Q_3(0)
		\end{bmatrix}
	 \begin{bmatrix}
		\lambda _1^{-1} & 0 & 0 \\
		0 & \lambda _2^{-1} & 0 \\
		0 & 0 & \lambda_3^{-1}
	\end{bmatrix} 
   	\begin{bmatrix}
   	Q_1(0) \\ Q_2(0) \\ Q_3(0)
   \end{bmatrix}
   \right) \\
   = \text{exp}\left( - \frac{D_{\text{eff}}}{2}Q^2_1(0) \lambda _1^{-1}\right)   \text{exp}\left( - \frac{D_{\text{eff}}}{2}Q^2_2(0) \lambda_2^{-1}\right)   \text{exp}\left( - \frac{D_{\text{eff}}}{2}Q^2_3(0) \lambda _3^{-1}\right)
	\end{multline}
with 
\begin{align} \label{eq:QfromG}
	Q_i(0)= \gamma \int_{0}^{t_f} \mathrm{d}t' \, e^{-\lambda_i t'} \, G'_i(t') \ , \qquad i=1,2,3 \ .
\end{align}

In the case when the $j$th eigenvalue $\lambda_j$ is small, so is $Q_j(0)$ due to the gradient echo condition, and the evaluation of the corresponding exponent on the right hand side of \eqref{eq:eugly} is numerically difficult. In this case, one can make use of the Taylor expansion of the exponential in \eqref{eq:QfromG} yielding
\begin{align}
\text{exp}\left( - \frac{D_{\text{eff}}}{2}Q^2_j(0) \lambda _j^{-1}\right) \approx \text{exp}\left( - \frac{D_{\text{eff}}}{2} \left[ \lambda_j \alpha_j^2 + \frac{1}{4} \lambda_j^3 \beta_j^2 -\lambda_j^2\alpha_j\beta_j \right]\right) \ ,
\end{align}
where
\begin{subequations}
\begin{align}
	\alpha_j &= \gamma \int_{0}^{t_f} \mathrm{d}\tau \, G'_j(\tau) \, \tau \\
	\beta_j &= \gamma \int_{0}^{t_f} \mathrm{d}\tau \, G'_j(\tau) \,\tau^2 \ .
\end{align}
\end{subequations}

\end{document}